\documentclass{naturemodified}
\usepackage{amsmath,amssymb,mathrsfs,latexsym,graphicx,psfrag}
\newcommand{\bs}[1]{\boldsymbol{#1}}

\title{{\small QUANTUM MANY BODY PHYSICS}\\
Confinement in a Quantum Magnet}
\author{Martin Greiter}

\begin{document}
\maketitle

\begin{affiliations}
\item[] Institut f\"ur Theorie der Kondensierten Materie,
  Karlsruhe Institute of Technology, 76128 Karlsruhe, Germany.  email:
  greiter@tkm.uni-karlsruhe.de
\end{affiliations}

\begin{abstract}
  The elementary excitations of a state of matter consisting of large
  collection of interacting particles can be very different from the
  original particles.  In the most interesting examples, the particles
  effectively decompose into smaller constituent particles, which only
  carry a fraction of their quantum numbers.  When these constituents
  are free, as in fractionally quantised Hall states, it is
  conceptually clear how to observe them.  But what if they are
  confined, as it might be the case in spin liquids hypothesised to
  describe high $\bs{T_c}$ superconductors?
\end{abstract}

The classic example of confinement among constituent particles is
given by the theory of the strong interactions in particle physics,
also known as quantum chromodynamics\cite{GreinerSchafer94} (QCD).
The theory assumes that hadrons, which most prominently include
protons, neutrons, and pions, consist of several smaller particles
called quarks, which among other quantum numbers carry fractional
charge and are held together through a non-Abelian gauge field
interactions, or equivalently, through the interchange of virtual
particles called gluons.  This interaction couples to an SU(3) quantum
number, which is called colour and can take the nominal values
``red'', ``blue'' or ``green''.  Unlike weak, electromagnetic, and
gravitational forces, the force mediated by the gluons does not
decrease with increasing distance, which confines the quarks into
bound states with no net colour, or technically speaking, SU(3) colour
singlets.  Depending on which quarks one combines to form these
singlets, one obtains protons, neutrons, and so on.  All the hadrons
are comparatively heavy, as a lot of energy is stored in the internal
field configuration where the quarks wiggle around each other
vigorously but cannot escape.
This theory has not led to quantitative advances in nuclear physics,
as contemporary methods do not allow us to evaluate systems consisting
of significant numbers of strongly interacting quarks, but has still
been confirmed experimentally, due to a property called ``asymptotic
freedom''.  This property implies that the quarks interact only weakly
if probed at sufficiently high energies, which makes them visible in
high-energy experiments.

In certain condensed matter systems, similar constituent particles
appear as collective excitations of strongly correlated many body
states.  The most established examples are the fractionally charged
quasiparticles in quantised Hall states\cite{Laughlin83prl1395} and
spinons---that is, particles with the spin of an electron but without
the charge, in models of antiferromagnetically interacting spins
(\emph{i.e.}, neighbouring spins like to point in opposite directions)
on a one-dimensional lattice\cite{Giamarchi04} (\emph{i.e.}, spin
chains).  In both these examples, the constituent particles are
``free'', \emph{i.e.}, deconfined, which greatly eased the task of
observing and hence establishing them.  There are other instances,
however, where theoretical models suggest that the constituent
particles are confined, as the quarks in QCD are.  It has been a long
outstanding problem to observe them experimentally in these systems,
as most probes just detect the bound states without revealing the
internal structure.  On page XXX of this issue, Lake and
colleagues\cite{Lake-0908.1038} report the first observation of a
crossover between confined and deconfined spinon excitations through
variation of the energy scale they employ in inelastic neutron
scattering experiments on a quantum magnet.  To borrow jargon from
particles physics, they observed ``asymptotically free'' spinons with
spin $1/2$ at high energy transfers and bound states of them with spin
$1$ at lower energies.  Or put more simply, they have observed
confinement in a condensed matter system.

To see how they accomplished this, a little bit of background is
helpful.  Let us begin with the fractional quantisation of charge in
the Laughlin state\cite{Laughlin83prl1395} describing the a quantised
Hall fluid at Landau level (LL) filling fraction 1/3.  There are three
times as many states in the lowest LL as there are electrons.
Analyticity properties require that the many body wave function has as
many zeros in each particle coordinate as there are states, that is,
three times as many as there are electrons.  The Fermi statistics of
the electrons requires that one of the zeros for a given electron
coordinate coincides with each of the positions of the other
electrons.  The distinguishing feature of the Laughlin state is that
all the remaining zeros are also attached to the electrons, such that
each coordinate coincides three zeros.  A true hole in the liquid is
consequently given by three zeros without an electron attached.  These
three zeros will repell each other, and each of them will form a
``quasihole''.  As the hole had an effective charge $+e$ (where $-e$
is the electron charge), each quasihole will have charge $+e/3$.  This
charge has been observed in resonant tunnelling
experiments\cite{Goldman-95s1010}.

The mechanism for fractional quantisation of spin in antiferromagnetic
spin chains\cite{Giamarchi04} is actually similar, except that the
electrons are now replaced by spin-flips, which carry spin one.  We
view the spin-up-flips as ``particles'' in a background where all the
spins point down.  Since the antiferromagnetic interaction favours
neighbouring spin to anti-align, it induces an effective repulsion
between the spin-flips, which takes the role of the Coulomb repulsion
between the electrons in the quantum Hall liquid. The emergence of
fractionally quantised excitations, ``spinons'' with one half of the
spin of a spin flip, is illustrated in Figure 1.  In analogy to the
creation of the hole in the quantum Hall fluid, we remove a
spin-up-flip or create a spin-down-flip by turning an up spin into a
down spin in Figure 1a.  This creates two domain walls (\emph{i.e.},
parallel spins) on each side of the spin we flipped, which propagate
independently, as shown in Figure 1b.  As the spin-flip carried spin
one, each of the domain walls or spinons will carry spin 1/2.

The most direct way to observe the spinons is through inelastic
neutron scattering, which in essence measures the energy absorption
spectrum for spin flips at various wave vectors.  If the spin flip
were to create only one particle, one would observe a resonance with
a well defined energy.  As it decays into two smaller constituent
particles, the spinons, there is a continuum of ways to distribute the
total momentum of the spin flip among the spinons.  This yields a
continuous absorption spectrum, and is exactly what Lake and
colleagues observe when probing their system at high energies.

Confinement among spinons results if one couples two antiferromagnetic
spin chains\cite{Dagotto-96s618,Shelton-96prb8521}, as illustrated in
Figure 1c.  If one creates two domain walls or spinons at some
distance along the chains, the spins on the rungs between the spinons
become ferromagnetically aligned (\emph{i.e.},  they point in the same
direction), which costs energy as the interaction across the rungs
favours them to align antiferromagnetically.  The associated cost in
energy is hence proportional to the number of these rungs, and the
force between the spinons does not decrease with the distance.

The energy gap associated with the confinement can now be understood
as the quantum mechanical zero-point energy of the constant force
oscillator describing the relative motion of the two
spinons\cite{greiter02prb054505}.  The first excited state of this
oscillator is also important, as the wave function describing it
is antisymmetric under spinon interchange while the ground state is
symmetric.  The wave function in spin space is therefore likewise
symmetric (\emph{i.e.}, a spin triplet) for the ground state and
antisymmetric (\emph{i.e.}, a spin singlet) for the first excited
state.

The energy gap for magnetic excitations in antiferromagnetic ladder
systems (\emph{i.e.}, coupled spin chains) is long
established\cite{Dagotto-96s618}.  But beyond theoretical
models\cite{Shelton-96prb8521,greiter02prb054505}, it is not at all
clear how one can establish, even as a matter of principle, that we
are really looking at confined spinons.  Lake and her colleagues
succeed through an ingenious combination of two factors.  First, the
material they probe is effectively critical at the relevant energies,
which implies that the lowest energy excitations are gapless.  Second,
by comparing the measured intensity of the magnetic absorption
spectrum around what would be the magnetic ordering wave vector if the
system were ordered to universal predictions of a conformal field
theory, the so-called SU(2) level $k=2S$ Wess-Zumino-Novikov-Witten
model\cite{witten84cmp455}, they are able to extract the spin of the
now effectively gapless excitations as they vary the energy at which
they measure the magnetic absorption spectrum.  This enables them to
establish that the spin of the critical low energy excitations is
effectively $S=1$ in the energy window between 10 and 32 meV, and
$S=1/2$ above a crossover regime extending up to roughly 70 meV.  In
other words, they observe how ``asymptotically free'' spinons at high
energies evolve into excitations with spin $S=1$ as they lower the
energy, and thereby show that the $S=1$ triplon excitations are bound
states of confined spinons with $S=1/2$ each.

Let me conclude with a personal view why this experiment is important.
One of the main problems in contemporary condensed matter physics, if
not physics in general, is the problem of high $T_{\rm c}$
superconductivity in the cuprates\cite{zaanen-06np138}.  The materials
consist of weakly coupled CuO layers, which are responsible for the
anomalous properties and for most purposes adequately described by
two-dimensional antiferromagnets doped with mobile holes.  We may view
the planes as infinite arrays of strongly coupled spin chains, as
compared to the weakly coupled pairs of chains investigated by Lake
\emph{et al.}.  Many of the key properties, including the
superconductivity and 
the anomalous properties of the so-called 'pseudo-gap' phase,
could be understood very plausibly if the holes where in fact
spinon-holon bound states held together by a strong confinement force.
If Lake and her colleagues could confirm this picture at a level
similar to the results reported for spinon confinement in coupled
chains, it would provide a huge step towards solving high
$T_{\rm c}$.


\newpage
\section*{Figure Captions}

{\bf Figure 1\ \ $\vert$\ \ Spinon confinement in coupled spin chains.}
{\bf a}, A spin flip in a spin chain results into two domain walls or
parallel spin on neighbouring sites. {\bf b}, These two domain walls or
spinons propagate independently and carry spin 1/2 each, since the
spin flip carried spin one. {\bf c}, When two spin chains are coupled
to form spin ladder, all the rungs in between the two spinons are
frustrated, which yields a linear confinement potential between them.
{\bf d}, The energy gaps for triplet and singlet excitations in the
ladder correspond to the ground state and first excited state energies
of the oscillator describing the relative motion of the spinons.  The
illustration in ({\bf a})-({\bf c}) is somewhat simplistic, as the
long range order present here is not present in the true ground
states, and the true spinons are not domain walls, but excitations of
spin 1/2 in an otherwise featureless spin liquid.

\vspace{80pt}
\begin{figure}[h]
  \begin{center}
    \includegraphics[width=0.97\linewidth]{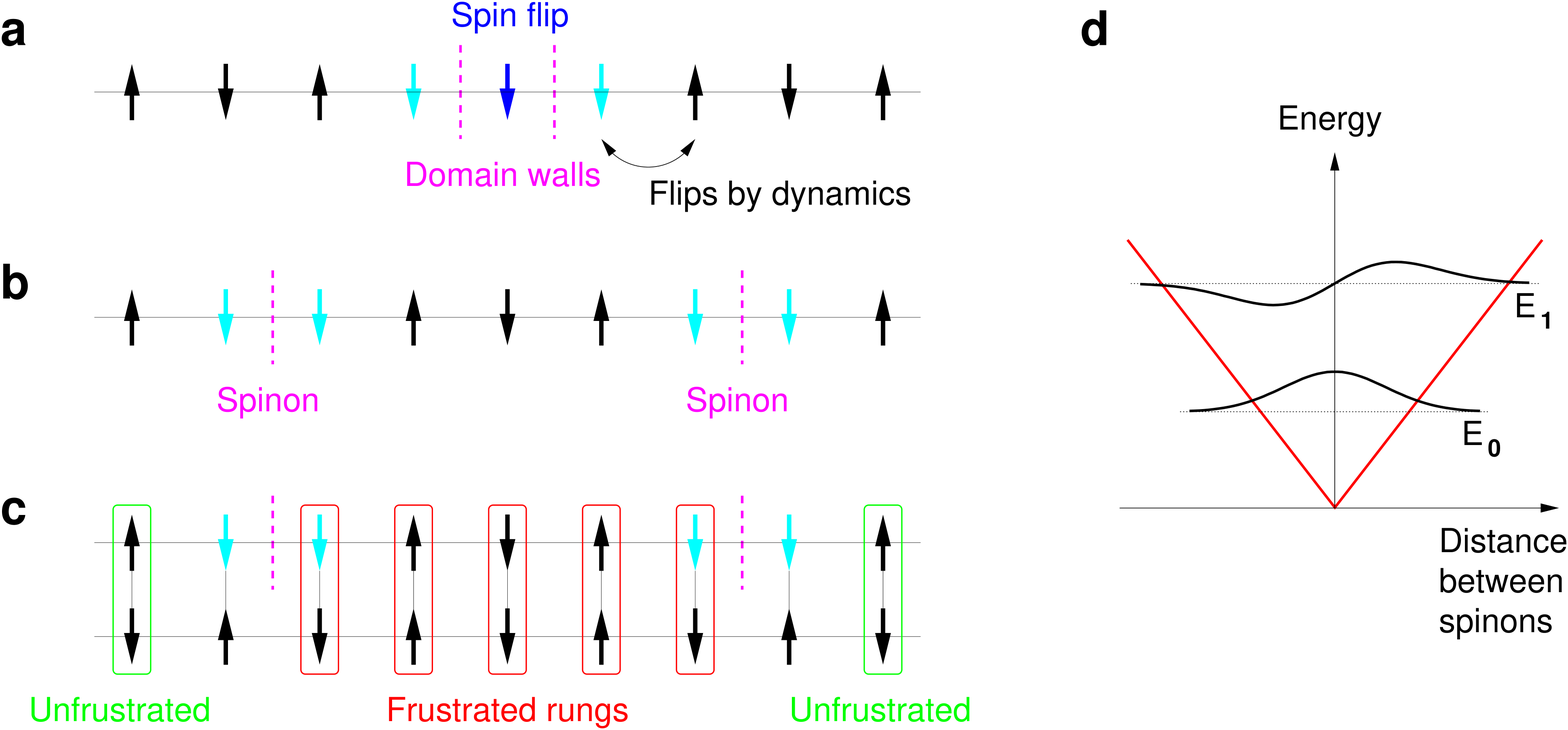}
  \end{center}
\end{figure}

\end{document}